\documentstyle[amssymb,preprint, aps]{revtex}

\begin{document}
\draft
\title{Avalanche dynamics in Bak-Sneppen evolution model observed with standard
distribution width of fitness}
\author{Chaohong Lee\thanks{%
Email address: chlee@wipm.whcnc.ac.cn}, Xiwen Zhu, and Kelin Gao}
\address{{\small \ }Laboratory of Magnetic Resonance and Atomic and Molecular
Physics, Wuhan Institute of Physics and Mathematics, The Chinese Academy of
Sciences, Wuhan, 430071, People's Republic of China}
\date{\today}
\maketitle

\begin{abstract}
We introduce the standard distribution width of fitness to characterize the
global and individual features of a ecosystem in the Bak-Sneppen evolution
model. Through tracking this quantity in evolution, a different hierarchy of
avalanche dynamics, $w_{0}$ avalanche is observed. The corresponding gap
equation and the self-organized threshold $w_{c}$ are obtained. The critical
exponents $\tau ,$ $\gamma \ $and $\rho $, which describe the behavior of
the avalanche size distribution, the average avalanche size and the
relaxation to attractor, respectively, are calculated with numerical
simulation. The exact master equation and $\gamma $ equation are derived.
And the scaling relations are established among the critical exponents of
this new avalanche.
\end{abstract}



\section{Introduction}

The biological evolution theory of Darwin is well known to all. The
biological evolution can be divided into two different types,
micro-evolution and macro-evolution, which focus on the dynamics of
evolution at the level of a single population and of all species in
ecosystem respectively. Up to now, many scaling laws have been found in
macroevolutionary data. The distribution $N(m)$ of extinction sizes for
families $m$ decreases with $m$ according to a power-law: $N(m)\varpropto
m^{-\alpha }$\cite{Raup1}. The distribution $N(t)$ of genera lifetimes $t$
follows $N(t)\varpropto t^{-k}$\cite{Raup2}. The distribution of the number
of genera $N(g)$ with $g$ species can be fitted with $N(g)\varpropto g^{-\mu
}$ \cite{Burlando}. The disappearance of species or genera is similar to
radioactive decay, in which the amount of the ''original element'' decays
exponentially with time. This phenomenon has been named as Red Queen effect%
\cite{Valen}. The extinction events exhibit punctuated patterns\cite{Gould}.
The punctuated equilibrium can be interpreted as a metastable state of the
dynamics of that single species\cite{Peliti}. If an ecosystem is in
equilibrium, every species is stable. However, if one of them fluctuates
with time, it will lead to the destabilization of its neighboring species.
Thus, the avalanche fluctuation will sweep through the evolution.

The existence of scaling laws in macro-evolution indicates that the dynamics
of large-scale evolution is the result of a self-organized critical process%
\cite{KSneppen,RVSole,PBak}. Based upon this view, Bak and Sneppen
introduced a toy model for species evolution\cite{Bak}. It is defined and
simulated as follows: (i) $L$ species are arranged on a one-dimensional
lattice with periodic boundary conditions. (ii) Random numbers $f_{i}$,
which are chosen from a uniform distribution between $0$ and $1$, $P(f)$,
are independently assigned to each site (species). (iii) At each time step,
the extremal site, i.e., the one with the smallest random number $f_{\min }$%
, and its nearest neighbors are mutated by assigning new random numbers in $%
P(f)$. As the system evolves it turns out that the species with fitness
significantly above the self-organized threshold $f_{c}$ (equal to $0.66702$%
\cite{PMB}) will never mutate, unless its neighbors mutate. After enough
mutations have occurred, the ecosystem reaches a statistically stationary
state in which the distribution for the fitness of all species is
statistically stationary.

Since the Bak-Sneppen model was defined, its punctuated patterns and
avalanche dynamics have been investigated extensively. Through observing the
fluctuations of the smallest fitness, the model shows punctuated equilibrium
behavior, i.e., it self-organizes into a critical state with intermittent
coevolutionary avalanches of all size\cite{Bak}. Paczuski, Maslov and Bak
studied the $f_{_{0}}$ avalanche (PMB avalanche), forward avalanche and
backward avalanche and derived the exact scaling equations, they also
established the relations among the critical exponents\cite{PMB}. Recently,
by monitoring the variations of the average fitness, the $\overline{f}_{0}$
avalanche (LC avalanche), is found, and the exact equations and scaling
relations are obtained\cite{LC1,LC2,LC3}. As we know, the avalanche dynamics
is a kind of macroscopic phenomenon in driven dissipative systems. The
detailed dynamics of these systems sensitively depend on the initial
configurations. But the distribution of avalanches, i.e., scaling law, does
not depend on such details due to the universality of complexity\cite
{Butenberg}. In the existing works, analyzing the smallest fitness only
directly connects with the feature of individuals, but it does not directly
represent the global feature; discussing the average fitness only directly
connects with the global dynamics, but it does not directly manifest the
difference in individuals. Can we find a new quantity, which can describe
the global feature and the individual difference simultaneously? In this
paper, we define such a quantity and observe the corresponding avalanche
dynamics with this quantity.

In the next section, the standard distribution width of fitness is defined,
and its gap equation is obtained. In Sec. III, the dynamics of avalanche is
analyzed in detail, the exact mast equation is derived, and the scaling
relations are established among the critical exponents. The last section is
a brief summary.

\section{Standard distribution width of fitness and its gap equation}

In the Bak-Sneppen model, the fitness of the $i$-th species is denote as $%
f_{i}$ $(i=1,2,\cdot \cdot \cdot ,L)$, with $L$ being the total number of
species. The fitness represents the population or the living capability of
the species, large fitness means immense population or great living
capability, and vice versa. To describe the average population or average
living capability of an ecosystem, Li and Cai defined the average fitness as%
\cite{LC1,LC2,LC3} 
\begin{equation}
f_{av}(s)=\frac{1}{L}\sum_{i=1}^{L}f_{i}(s).
\end{equation}
where $f_{i}(s)$ is the fitness of the $i$-th species at time $s$. Based
upon the average fitness, we define the standard distribution width of
fitness as 
\begin{equation}
w(s)=\sqrt{\frac{1}{L}\sum_{i=1}^{L}(f_{i}(s)-f_{av}(s))^{2}}.
\end{equation}
From the above definition, we know that the standard distribution width of
fitness simultaneously manifests both global dynamics and individual
difference. At the beginning of the evolution, i.e., $s=0$, the fitnesses
are uniformly distributed in $(0,1)$. Thus for an infinite-size ecosystem $%
(L\rightarrow \infty )$, the initial average fitness $f_{av}(0)$ equals $0.5$%
, the initial standard distribution width $w(0)$ is $\sqrt{3}/6$. But for a
finite-size ecosystem, $f_{av}(0)$ and $w(0)$ fluctuate around the above
values due to the finite size. Because of the absence of mutations or
updates, $w(0)$ does not show the correlation between species, only when the
evolution goes on, such correlation is contained in $w(s)$ gradually.

As time increases, $f_{av}(s)$ shows an increasing tendency\cite{LC1},
whereas, $w(s)$ shows a decreasing tendency (see the bottom of Fig. 1).
After a long transience, i.e., $s\rightarrow \infty $, the ecosystem
self-organizes to a critical state, and, $f_{av}(s)$ and $w(s)$ approach to
different thresholds. When $s\rightarrow \infty ,L\rightarrow \infty $,
almost all species have fitnesses above the threshold $f_{c}=0.66702$\cite
{PMB}, and the fitnesses are uniformly distributed in $(f_{c},1)$. With the
definitions of $f_{av}(s)$ and $w(s)$, we obtain the relation among the
thresholds 
\begin{equation}
w_{c}=\frac{1}{2\sqrt{3}}(1-f_{c})=\frac{1}{\sqrt{3}}(1-f_{avc}).
\end{equation}
Here, $w_{c}$ and $f_{avc}$ are thresholds of $w(s)$ and $f_{av}(s)$
respectively. Thus, one gets $w_{c}=0.09623$ and $f_{avc}=0.83351$. For the
finite-size ecosystem, the self-organized critical thresholds will fluctuate
around the above values, and this fluctuation decreases with the increase of 
$L$.

Similar to the definition in Refs.\cite{PMB,LC1}, we define the gap of $w(s)$
as its envelope: 
\begin{equation}
W(s)=\left\{ 
\begin{array}{lll}
w(0) & , & s=0, \\ 
\min (w(s\acute{})|s\acute{}\in [0,s]) & , & s>0.
\end{array}
\right.
\end{equation}
The above definition means that the current gap of $w(s)$ is the most
minimum of all $w(s\acute{})$ $(0\leq s\acute{}\leq s)$. In the top of Fig.
1, we show the gap $W(s)$ as a function of $s$. With the increase of time, $%
W(s)$ gradually approaches close to the threshold. By definition, when the
gap falls to its next lower value, the separate instances are separated by
avalanches. The avalanches correspond to plateaus in $W(s)$ during which $%
w(s)>W(s)$. A new avalanche is initiated when the gap falls, and the old
avalanche terminates at the same time. Once a avalanche is over it will
never affect the behavior of any subsequent avalanche. When the gap
decreases, the probability for $w(s)$ to jump above $W(s)$ increases, thus
longer and longer avalanches happen typically. Finally, the ecosystem falls
into statistically stationary state.

Following Refs.\cite{PMB,LC1}, we derive the gap equation of $W(s)$. If the
gap is changed from $W(s)$ to $W(s)+\Delta W$, the average number of
avalanches occurred is $N_{av}=-\Delta WL/(W(s)-w_{c})$. Thus, when $%
L\rightarrow \infty $, the average number of time steps required for
increasing the gap from $W(s)$ to $W(s)+\Delta W$ is $\Delta s=\left\langle
S\right\rangle _{W(s)}N_{av}=-\left\langle S\right\rangle _{W(s)}\Delta
WL/(W(s)-w_{c})$, where $\left\langle S\right\rangle _{W(s)}$ is the average
size of avalanche of the current plateau of the gap. So the differential
equation for $W(s)$ is 
\begin{equation}
\frac{dW(s)}{ds}=\lim_{\Delta W,\Delta s\rightarrow 0}\frac{\Delta W}{\Delta
s}=-\frac{W(s)-w_{c}}{\left\langle S\right\rangle _{W(s)}L}.
\end{equation}
Eq.(5) describes the relaxation to attractor. If the law of $\left\langle
S\right\rangle _{W(s)}$ is obtained, one can derive the law of relaxation to
attractor from the above equation.

\section{Avalanche dynamics}

We know that all self-organized critical systems will exhibit a power-law
avalanche dynamics. This has been confirmed by a lot of models with
self-organized criticality\cite
{Bak,PMB,LC1,LC2,LC3,Jensen,Flyvbjerg,Maslov1,Maslov2,Lowen,Christensen,Datta}%
. In this section, based on the definition of PMB avalanche\cite{PMB} and LC
avalanche\cite{LC1}, the definition of a new avalanche, $w_{0\text{ }}$%
avalanche, is presented. Then its critical exponents are calculated with
numerical simulation and its exact master equation is derived. Lastly, the
scaling relations are established.

The definition for $w_{0\text{ }}$avalanche is as follows. Similar to those
used in Refs.\cite{PMB,LC1}, we introduce an auxiliary parameter $w_{0}$,
where $w(0)>w_{0}>w_{c}$. Suppose that at time step $s_{1}$, the current
standard distribution width $w(s_{1})$ is less than $w_{0}$. If the next
standard distribution width $w(s_{1}+1)$ is larger than $w_{0}$, a
creation-annihilation branching process is initiated. The avalanche will
continue to run unless $w(s)$ becomes less than $w_{0}$. This means that, at
time $s$, if all $w(s\acute{})>w_{0}$ for $(s_{1}<s\acute{}<s-1)$, the
current avalanche continues. In terms of the above definition, the size of
an avalanche is the number of time steps of subsequent punctuations above $%
w_{0}$. If the first appearance of $w(s)<w_{0}$ after time $s_{1}$ occurs at
time $s_{1}+S$, the size of the current avalanche is $S$.

Clearly, the above definition ensures the hierarchical structure, i.e.,
larger avalanches consist of smaller ones. With the decreasing of $w_{0}$,
smaller avalanches combine into larger ones; if $w_{0}$ is set as $w_{c}$,
the infinite-size avalanche appears. In the other case, with the increasing
of $w_{0}$, larger avalanches split into smaller ones. Thus, the cutoff
effects is unavoidable when $w_{0}$ is not chosen as $w_{c}$. Nevertheless,
the same scaling laws can be obtained when $w_{0}$ is very close to the
threshold $w_{c}$. For the Bak-Sneppen model with $200$ species, choosing $%
w_{0}$ as $0.1100$, we find by numerical simulation that the distribution of
avalanche size follows the power law 
\begin{equation}
P(S)\varpropto S^{-\tau },
\end{equation}
where $P(S)$ is the probability of avalanche of size $S$, $\tau $ equals $%
1.63\pm 0.06$ (see Fig. 2). Because of the cutoff effects, the average
avalanche size varies when $w_{0}$ changes. Similar to those in Refs.\cite
{PMB,LC2,LC3}, our numerical results show that the average size of $w_{0}$
avalanche obeys 
\begin{equation}
\left\langle S\right\rangle _{w_{0}}\varpropto (w_{0}-w_{c})^{-\gamma }.
\end{equation}
Here, $w_{c}$ is the threshold, the exponent $\gamma $ equals $2.57\pm 0.05$
(see Fig. 3). This law means that the divergence of the average avalanche
size satisfies power law. Our numerical simulation also shows that the
relaxation to attractor abides 
\begin{equation}
(W-w_{c})\varpropto s^{-\rho }.
\end{equation}
Here, $s$ is the time steps, the exponent $\rho $ equals $0.39\pm 0.03$.
Obviously, the value of $\rho $ is very close to $1/\gamma $.

So far, we have found some scaling laws of the $w_{0\text{ }}$avalanche by
numerically simulations. Now, we will give some exact results. To describe
the cascade process of smaller avalanches combining into larger ones when
the parameter $w_{0}$ is changed, an exact master equation is derived as
follows. Denoting the probability of $w_{0\text{ }}$avalanche of size $S$ as 
$P(S,w_{0})$, when parameter changes from $w_{0}$ into $w_{0}+\Delta w_{0}$,
some $w_{0\text{ }}$avalanches merge to larger $w_{0}+\Delta w_{0}$
avalanches, and some small avalanches merge into $w_{0\text{ }}$avalanche.
Thus, at the same time, the probability of a avalanche with a given size
flows in and out. Because the termination of $w_{0\text{ }}$avalanche is
uncorrelated, the probability of a $w_{0\text{ }}$avalanche merging to $%
w_{0}+\Delta w_{0}$ avalanche is proportional to $-\Delta
w_{0}/(w_{0}-w_{c}) $\cite{LC1,LC2}. Thus the probability which flows out
from $P(S,w_{0})$ can be written as 
\begin{equation}
\Delta _{out}P=-\lambda P(S,w_{0})\Delta w_{0}/(w_{0}-w_{c}),
\end{equation}
where $\lambda $ is a constant. The probability which flows in $P(S,w_{0})$
is 
\begin{equation}
\Delta _{in}P=-\lambda
\sum\limits_{S_{1}=1}^{S-1}P(S_{1},w_{0})P(S-S_{1},w_{0})\Delta
w_{0}/(w_{0}-w_{c}).
\end{equation}
When $w_{0}$ approaches $w_{c}$ and $\Delta w_{0}$ approaches zero, we
obtain the following master equation 
\begin{equation}
-(w_{0}-w_{c})\frac{\partial P(S,w_{0})}{\partial w_{0}}=-\lambda
P(S,w_{0})+\lambda \sum\limits_{S_{1}=1}^{S-1}P(S_{1},w_{0})P(S-S_{1},w_{0}).
\end{equation}
As pointed out in the previous section, with the increasing of time, the
standard distribution width shows a decreasing tendency. So, when $w_{0}$
decreases, the first term on the right side of the master equation reflects
the loss of $S$-size avalanche, and the second term depicts the gain of $S$%
-size avalanche.

Similar to Ref.\cite{LC2}, after defining a new variable $u=-\ln
(w_{0}-w_{c})$, the master equation becomes 
\begin{equation}
\frac{\partial P(S,u)}{\partial u}=-\lambda P(S,u)+\lambda
\sum\limits_{S_{1}=1}^{S-1}P(S_{1},u)P(S-S_{1},u).
\end{equation}
Through performing a Laplacian transformation for $P(S,u)$, i.e., 
\begin{equation}
p(\alpha ,u)=\sum\limits_{S=1}^{\infty }P(S,u)e^{-\alpha S},
\end{equation}
thus the master equation reads as 
\begin{equation}
\frac{\partial p(\alpha ,u)}{\partial u}=\lambda p(\alpha ,u)[p(\alpha
,u)-1].
\end{equation}
When $\alpha =0$, $p(\alpha ,u)=\sum\limits_{S=1}^{\infty }P(S,u)e^{-\alpha
S}=1$, this is the normalization of $P(S,u)$. While $\alpha >0$, one gets $%
0<p(\alpha ,u)<1$. Expanding both sides of Eq.(14) as Taylor series
throughout the neighborhood of the point $\alpha =0$, one can obtain 
\begin{eqnarray}
&&\frac{\partial }{\partial u}[\alpha \left\langle S\right\rangle _{u}-\frac{%
1}{2!}\alpha ^{2}\left\langle S^{2}\right\rangle _{u}+\frac{1}{3!}\alpha
^{3}\left\langle S^{3}\right\rangle _{u}\cdot \cdot \cdot ]  \nonumber \\
&=&\lambda [1-\alpha \left\langle S\right\rangle _{u}+\frac{1}{2!}\alpha
^{2}\left\langle S^{2}\right\rangle _{u}-\frac{1}{3!}\alpha ^{3}\left\langle
S^{3}\right\rangle _{u}\cdot \cdot \cdot ]  \nonumber \\
&&\times [\alpha \left\langle S\right\rangle _{u}-\frac{1}{2!}\alpha
^{2}\left\langle S^{2}\right\rangle _{u}+\frac{1}{3!}\alpha ^{3}\left\langle
S^{3}\right\rangle _{u}\cdot \cdot \cdot ],
\end{eqnarray}
where $\left\langle S^{k}\right\rangle _{u}=\sum\limits_{S=1}^{\infty
}S^{k}P(S,u)$. Comparing the coefficients of different powers of $\alpha $
in Eq.(15), one gains an infinite series of exact equations. The first exact
equations can be expressed as 
\begin{equation}
\frac{\partial }{\partial u}\ln \left\langle S\right\rangle _{u}=\lambda .
\end{equation}
Replacing $u$ as $-\ln (w_{0}-w_{c})$, we obtain the $\gamma $ equation 
\begin{equation}
\frac{\partial }{\partial w_{0}}\ln \left\langle S\right\rangle _{w_{0}}=-%
\frac{\lambda }{w_{0}-w_{c}}.
\end{equation}
Noting the scaling law $\left\langle S\right\rangle _{w_{0}}\varpropto
(w_{0}-w_{c})^{-\gamma }$, inserting it into the above equation, one can
easily gain 
\begin{equation}
\gamma =\lambda .
\end{equation}

By far, we have obtained three critical exponents of the $w_{0}$ avalanche: $%
\tau $, $\gamma \ $and $\rho $, which describe the behavior of the avalanche
size distribution, the average avalanche size, and the relaxation to
attractor, respectively. As listed in Ref.\cite{PMB}, there are three other
critical exponents: $D$, $\sigma $ and $\upsilon $. The avalanche dimension $%
D$ is defined by the following scaling relation 
\begin{equation}
S\varpropto R^{D}.
\end{equation}
where $S$ is the avalanche size (temporal duration), $R$ is the spatial
extent. The $\sigma $ and $\upsilon $ are defined by the following scaling
laws of cutoff 
\begin{equation}
S_{co}\varpropto (w_{0}-w_{c})^{-1/\sigma },
\end{equation}
\begin{equation}
R_{co}\varpropto (w_{0}-w_{c})^{-\upsilon }.
\end{equation}
The symbols $S_{co}$ and $R_{co}$ are the cutoff of the avalanche size and
that of spatial extent of the avalanche.

Although there are six critical exponents, only part of them are
independent. After choosing $\tau $ and $\gamma $ as the independent
exponents, the other ones can be expressed as functions of them. Inserting
the scaling law $\left\langle S\right\rangle _{w_{0}}\varpropto
(w_{0}-w_{c})^{-\gamma }$ into the gap equation (Eq.(5)), and noting the
scaling law $(w_{0}-w_{c})\varpropto s^{-\rho }$, one can find 
\begin{equation}
\rho =\gamma ^{-1}.
\end{equation}
Integrating the equation $\left\langle S\right\rangle
_{w_{0}}=\int_{1}^{S_{co}}SP(S,w_{0})dS$, it is easy to obtain 
\begin{equation}
\left\langle S\right\rangle
_{w_{0}}=\int_{1}^{S_{co}}SP(S,w_{0})dS=\int_{1}^{S_{co}}S^{1-\tau
}dS\varpropto S_{co}^{2-\tau }\varpropto (w_{0}-w_{c})^{-\frac{2-\tau }{%
\sigma }}.
\end{equation}
Comparing with $\left\langle S\right\rangle _{w_{0}}\varpropto
(w_{0}-w_{c})^{-\gamma }$, one can get 
\begin{equation}
\sigma =\frac{2-\tau }{\gamma }.
\end{equation}
Noting that the average number of the covered sites $\left\langle
n_{cov}\right\rangle _{w_{0}}$ covered by a $w_{0}$ avalanche scales near
the critical point as\cite{PMB} $\left\langle n_{cov}\right\rangle
_{w_{0}}\varpropto R_{cov}\varpropto R\varpropto (w_{0}-w_{c})^{-1}$and $%
S\varpropto R^{D}$, so $\left\langle n_{cov}\right\rangle _{w_{0}}\varpropto
S^{1/D}$, integrating the equation $\left\langle n_{cov}(S)\right\rangle
_{w_{0}}=\int_{1}^{S_{co}}n_{cov}(S)P(S,w_{0})dS$ gives 
\begin{equation}
\left\langle n_{cov}(S)\right\rangle _{w_{0}}\varpropto
\int_{1}^{S_{co}}S^{1/D-\tau }dS\varpropto (w_{0}-w_{c})^{-\frac{1-\tau +1/D%
}{\sigma }}.
\end{equation}
Thus, the avalanche dimension $D$ can be written as 
\begin{equation}
D=\frac{1}{\tau +\sigma -1}=\frac{\gamma }{\gamma (\tau -1)+2-\tau }.
\end{equation}
Noting $S\varpropto R^{D}$ and $R_{co}\varpropto (w_{0}-w_{c})^{-\upsilon }$%
, one can immediately obtain 
\begin{equation}
S_{co}\varpropto R_{co}^{D}\varpropto (w_{0}-w_{c})^{-\upsilon D}.
\end{equation}
Comparing with $S_{co}\varpropto (w_{0}-w_{c})^{-1/\sigma }$, it is easy to
know 
\begin{equation}
\upsilon =\frac{1}{\sigma D}=1+\frac{\gamma (\tau -1)}{2-\tau }.
\end{equation}
Up to now, the scaling relations among the six critical exponents have been
successfully established. The existence of two independent critical
exponents means that there are two kinds uncorrelated critical behavior in $%
w_{0}$ avalanche.

\section{Summary}

In conclusion, by defining a new quantity, the standard distribution width
of fitness, a new avalanche, $w_{0}$ avalanche, is observed in the evolution
of Bak-Sneppen model. We declare this avalanche as a different hierarchy of
avalanche, because firstly, the observed quantity is different from those of
Refs.\cite{Bak,PMB} and Refs.\cite{LC1,LC2,LC3}; and secondly, the values of
the critical exponents are also different from those found in Refs.\cite
{Bak,PMB} and Refs.\cite{LC1,LC2,LC3}.

From the definition of the standard distribution width of fitness, we easily
obtain the corresponding gap equation and the self-organized threshold.
Apparently, with this definition, the global feature and the difference
among individuals are simultaneously described by the standard distribution
width. So, for the evolutionary distribution with changeless or small
changed average values, the standard distribution width is a good quantity
to observe the evolutionary dynamics. According to the definition of PMB
avalanche and LC avalanche, the definition of the $w_{0}$ avalanche is
presented. With numerical simulation, the critical exponents $\tau $, $%
\gamma \ $and $\rho $ are obtained. Then the exact master equation is
derived. From the master equation, the $\gamma $ equation is immediately
obtained. Finally, combining all scaling laws, the scaling relations are
successfully established among the critical exponents. With these relations,
we find there are only two uncorrelated critical behavior in the $w_{0}$
avalanche, for there exist only two independent critical exponents.

\begin{center}
{\large Acknowledgment}
\end{center}

{The work is supported by the National Natural Science Foundations of China
and Foundations of Chinese Academy of Sciences.} The authors are very
grateful the help of Prof. X. Cai and Dr. W. Li of Institute of Particle
Physics in Hua-zhong Normal University. 
\begin{figure}[tbp]
\caption{The standard distribution width and its gap. The above row
corresponds to the variation of the gap $W(s)$, the bottom row corresponds
to the fluctuation of $w(s)$.}
\end{figure}
\begin{figure}[tbp]
\caption{The power-law of the avalanche size distribution. }
\end{figure}
\begin{figure}[tbp]
\caption{The power-law of the divergence of the average avalanche size.}
\end{figure}

\end{document}